\documentclass[aps,pra,amsmath,amssymb,reprint,superscriptaddress]{revtex4-2}
\usepackage{xcolor,mathtools,graphicx,dcolumn,bm,hyperref}
\usepackage{cleveref}

\crefname{figure}{Fig.}{Figs.}
\Crefname{figure}{Fig.}{Figs.}
\begin{document}
\title{Exploring transition pathways in the Landau--Brazovskii model}

\author{Zhiyi Zhang}
 \affiliation{School of Mathematical Sciences, Peking University, Beijing 100871, China}
 
\author{Gang Cui}
 \affiliation{Hunan Key Laboratory for Computation and Simulation in Science and Engineering, Key Laboratory of Intelligent Computing and Information Processing of Ministry of Education, School of Mathematics and Computational Science, Xiangtan University, Xiangtan, Hunan 411105, China}
 
\author{Kai Jiang}
 \affiliation{Hunan Key Laboratory for Computation and Simulation in Science and Engineering, Key Laboratory of Intelligent Computing and Information Processing of Ministry of Education, School of Mathematics and Computational Science, Xiangtan University, Xiangtan, Hunan 411105, China}
 
\author{An-Chang Shi}
 \affiliation{Department of Physics and Astronomy, McMaster University, Hamilton, Ontario L8S 4M1, Canada}
 
\author{Pingwen Zhang}
 \email{pzhang@pku.edu.cn}
 \affiliation{School of Mathematics and Statistics, Wuhan University, Wuhan 430072, China}
 \affiliation{School of Mathematical Sciences, Peking University, Beijing 100871, China}
 
\author{Jianyuan Yin}
 \email{jyyin@bnu.edu.cn}
 \affiliation{School of Mathematical Sciences, 
 Laboratory of Mathematics and Complex Systems, Ministry of Education, 
 Beijing Normal University, Beijing 100875, China}
 
\author{Lei Zhang}
 \email{zhangl@math.pku.edu.cn}
 \affiliation{Beijing International Center for Mathematical Research, 
 Center for Quantitative Biology, 
 Center for Machine Learning Research, 
 Peking University, Beijing 100871, China}

\date{\today}

\begin{abstract}
The Landau--Brazovskii model provides a theoretical framework for describing various phases arising from competing short- and long-range interactions in many physical systems.
In this work, we investigate phase transitions among various ordered phases within the three-dimensional Landau--Brazovskii model.
We construct the phase diagram of this model, which encompasses eight distinct phases, and systematically compute the transition pathways connecting various metastable and stable states using the Landau--Brazovskii saddle dynamics. 
Along each transition pathway, the critical nucleus is identified with some detailed analyses of its shape, energy barrier, and Hessian eigenvalues. 
Furthermore, we explore how the transition state is influenced by model parameters, revealing systematic trends in critical nucleus sizes and energy barrier heights. 
Our results provide a comprehensive characterization of the nucleation mechanisms within the Landau--Brazovskii model and offer valuable insights into the structural transformations of modulated-phase systems.
\end{abstract}

\maketitle

\section{Introduction}

The Landau--Brazovskii (LB) model is a prototypical continuum theory for systems that undergo ordering at a finite wavenumber, producing spatially modulated phases with a characteristic length scale. 
This model, based on Landau theory, was originally proposed by Brazovskii as a general framework to describe isotropic-anisotropic phase transitions \cite{brazovskii1975phase}, and has served as a generic model for the study of modulated phases in diverse physical systems, such as magnetic systems, fluid mechanics, block copolymers, and even nuclear pasta in neutron stars \cite{andelman2009modulated,caplan2017colloquium}. 
This model has also been applied to general systems of hard spheres with additional pairwise interactions that are short-ranged in the real space or strongly localized around a finite wavenumber in the reciprocal space \cite{ciach2013origin}. 
Closely related to the Swift--Hohenberg (SH) equation \cite{sw} and Landau theory near a Lifshitz point \cite{goldenfeld2018lectures}, the LB free-energy framework has become a standard variational setting for rigorous analysis of periodic minimizers, energy scaling, pattern selection, and metastability \cite{peletier2012spatial}.

Within this framework, ordering at a preferred finite wavenumber naturally gives rise to a family of spatially periodic phases. 
Within the mean-field approximation, it has been shown that the LB theory predicts various ordered structures, including the lamellar (LAM), hexagonally-packed cylinder (HEX), and body-centered cubic (BCC) phases \cite{lb2}. 
The phase diagram was constructed by comparing the free energies of the disordered phase, LAM, HEX, BCC and double gyroid (DG) phases  \cite{AClbmodel}, and additional crystalline structures, such as the face-centered cubic (FCC) phase and the Frank--Kasper phases including A15 and $\sigma$, were also identified within the LB model \cite{diagram}. 
Extensive studies of these ordered structures have also been carried out experimentally in block-copolymer systems \cite{copolymerexp1, copolymerexp2,copolymerexp4}.

Beyond characterizing ordered phases, the LB model has also been widely used to investigate phase transitions between metastable and stable phases. 
Within this framework \cite{lb1,lb3,lb4}, order-to-order transformations such as the HEX--LAM and BCC--HEX transitions have been analyzed, and related studies of cylinder--gyroid interfacial systems have also been performed \cite{yao2022transition}. 
Experiments on block copolymers have shown that these transitions proceed via structural rearrangements of macromolecular aggregates between distinct periodic morphologies.
These transitions can proceed epitaxially, whereby features of the product phase emerge directly from the parent phase without long-range material transport \cite{copolymerepitaxi,copolymer2}. 
Such epitaxial behavior has also been reproduced within the LB model, exhibiting its ability to capture order-to-order transition pathways.

As theoretical approaches continue to advance our understanding of phase transitions, determining the transition pathways between different minima, which is characterized by the minimum energy paths (MEPs), has been a central focus of the research. 
An MEP connects local free-energy minima with transition states, which correspond to index-1 saddle points, and along the pathway the gradient of the free energy is parallel to the tangent direction.
Such a pathway provides essential physical insight into the nucleation mechanism, the associated energy barriers, and the structural evolution during phase transformations. 
Compared with the numerical identification of local minima, which correspond to stable or metastable states, locating transition states is considerably more challenging because of their inherent instability.
Numerous numerical methods have been proposed to study phase transition and calculate the transition pathway, which can be primarily classified into two categories: path-finding methods and surface-walking approaches. 
The path-finding methods, such as nudged elastic band methods \cite{neb1} and string methods \cite{string1,scft3}, require knowledge of both initial and final states and a suitable initialization connecting them. 
The latter approaches start from a single state and locate the transition state directly without \textit{a priori} knowledge of the final state. 
Representative methods include gentlest ascent dynamics \cite{gad1}, dimer methods \cite{henkelman1999dimer, zhang2016optimization}, the activation-relaxation technique \cite{cances2009some}, and the eigenvector-following methods \cite{cerjan1981finding}.

On a complex energy landscape with multiple local minima, numerous transition states may arise, and the MEP with the lowest energy barrier corresponds to the most probable transition process. 
In this LB model, the high dimensionality of the free-energy landscape and the localized nature of critical nuclei further exacerbate the computational challenges of identifying this optimal MEP. 
Furthermore, degeneracies in the Hessians at ordered phases prevent the surface-walking approaches for transition states from climbing out of the basin.
Recently, the solution landscape method \cite{PRLhisd,yin2021searching}, which constructs a pathway map on the energy landscape by systematically calculating saddle points with high-index saddle dynamics methods \cite{HiOSD,yin2021searching}, has effectively addressed these challenges in complex energy landscapes with multiple local minima across diverse physical systems, including liquid crystals, quasicrystals, and Bose--Einstein condensates \cite{PRLhisd,yinOnsager,PNAShisd,yinBose}.
Motivated by this, we develop an LB saddle dynamics (LBSD) method to efficiently identify the desired transition pathways and gain deeper insights into modulated-phase systems.

In this paper, we first introduce the LB free-energy functional as a mathematical model for modulated-phase systems and the LBSD method employed in our computations for phase transitions. 
Then, we construct the phase diagram and calculate transition pathways connecting different stable and metastable phases under varying parameter conditions.
Along each transition pathway, we identify the critical nucleus and its associated energy barrier, while further analyzing the nucleus shape and the Hessian eigenvalues of the saddle points. 
We further explore how the model parameters influence the size of the critical nuclei and the height of the energy barriers.

\section{Landau--Brazovskii model}

In general, the LB model can be stated as \cite{AClbmodel}:
\begin{equation}\label{eq:lb0}
\begin{aligned}
\mathcal{E}[\phi] = \int & \left\{ \frac{\xi_0^2}{8q_0^2} \left| (\nabla^2 + q_0^2)\phi \right|^2 + \frac{\tau}{2}\phi^2\right.\\
&\left. - \frac{\gamma}{3!}\phi^3 + \frac{\lambda}{4!}\phi^4 \right\}\mathrm{d}\mathbf{r}
\end{aligned}
\end{equation}
where $\phi(\mathbf{r})$ is the scalar order parameter defined on the three-dimensional space, $\tau$ is the reduced temperature, $q_0$ is the critical wavelength, $\xi_0$ is the bare correlation length, $\gamma$ is related to the asymmetry, and $\lambda > 0$ is a phenomenological constant. 
Furthermore, the order parameter $\phi$ satisfies the mass-conservation constraint $\int \phi \; \mathrm{d} \mathbf{r} = 0$. 
Note that $\mathcal{E}$ is invariant under the transformation $(\phi,\gamma)\to(-\phi,-\gamma)$, so only $\gamma \geqslant 0 $ is considered.

The LB model \eqref{eq:lb0} currently involves five scalar parameters $\xi_0, q_0, \tau, \gamma,$ and $\lambda$. 
By applying $1/q_0$ as the length scale ($\tilde{\mathbf r} = q_0 \mathbf r$), the rescaled quantities can be defined as
\begin{equation*}
\tilde{\xi} = \dfrac{q_0^2 \xi_0^2 }{4\lambda}, \; \; \;
\tilde{\mathcal{E}} = \dfrac{q_0^3 \mathcal{E}}{\lambda \tilde{\xi}^4 }, \; \; \;
\tilde{\tau} = \dfrac{\tau }{\lambda \tilde{\xi}^2}, \; \; \;
\tilde{\gamma} = \dfrac{\gamma }{\lambda\tilde{\xi}}, \; \; \;
\tilde{\phi} = \dfrac{\phi}{\tilde{\xi}}.
\end{equation*}
This rescaling process implies that the properties of the LB model can be determined by only two scalar parameters $\tilde{\tau}$ and $\tilde{\gamma}$. 
Therefore, the LB free-energy model can be rewritten in this following form (tilde omitted):
\begin{equation}\label{eq:lb}
\mathcal{E}[\phi] = \int  \left\{ \frac{1}{2} \left| (\nabla^2 + 1) \phi \right|^2+ \frac{\tau}{2}\phi^2\
 - \frac{\gamma}{3!}\phi^3 + \frac{1}{4!}\phi^4 \right\}\mathrm{d}\mathbf{r},
\end{equation}
with a mass-conservation constraint
\begin{equation}\label{eq:lbcons}
\int \phi \;\mathrm{d}\mathbf{r} =0. 
\end{equation}
In this manuscript, we employ this form of the LB free-energy model to consider multiple three-dimensional phases and their phase transitions.

By including a cubic term that characterizes the asymmetry between ordered phases, the LB free-energy framework provides a natural variational setting for describing metastable states and first-order order-to-order transitions, which are central to the present study \cite{AClbmodel,diagram}. 
The LB model \eqref{eq:lb} is closely related to the SH model \cite{sw}, which is also a canonical model for pattern formation arising from finite-wavenumber instabilities. 
The SH equation can typically be interpreted as an $L^2$ gradient flow associated with an LB-type free-energy functional, in contrast to the mass-conserved dynamics often considered in the LB setting.
The classical SH model has also been generalized to include cubic or higher-order terms in the energy functional for the study of pattern formation \cite{burke2006localized,houghton2011swift}. 
The Ohta--Kawasaki (OK) model represents another important class of models dominated by a finite-wavelength instability for describing microphase separation in block-copolymer systems \cite{ohta1986equilibrium,nishiura1995some}.
The OK model arises from the competition between short-range interfacial energy and long-range interactions, where the latter is represented by an explicit nonlocal term originating from polymer chain connectivity.
Consequently, the OK model is also capable of producing multiple three-dimensional ordered morphologies qualitatively similar to those of the LB model---including LAM, DG, HEX, BCC, and FCC phases \cite{dawson2021ginzburg}.

For simplicity, we denote a linear operator by $\mathcal{L}$,
\begin{equation*}
    \mathcal{L}\phi=(\nabla^2 + 1)^2 \phi+\tau\phi, \nonumber
\end{equation*}
and the last two nonlinear terms in \eqref{eq:lb} (and their negative derivatives) by
\begin{equation*}
    h(\phi) = -\frac{\gamma}{3!} \phi^3 + \frac{1}{4!}\phi^4, \quad 
    g(\phi) = -h'(\phi) = \frac{\gamma}{2} \phi^2  - \frac{1}{3!}\phi^3. \nonumber
\end{equation*}

In general, the scalar order parameter $\phi$ is defined on the entire three-dimensional space $\mathbb{R}^3$. 
To enable numerical implementations, we utilize the crystalline approximant method (CAM) \cite{cam2,jiang2025approximation} and approximate different phases with periodic crystal structures on finite domains in the three-dimensional space.
Specifically, we consider the order parameter $\phi$ on a box domain $\Omega$ with periodic boundary conditions to characterize each crystalline phases.
The subsequent discussion focuses on the free-energy density of the LB model integrated on the domain $\Omega$, 
\begin{equation}\label{eq:lbwomega}
f[\phi] =  \frac{1}{|\Omega|}\int_\Omega \left[ \frac{1}{2} \left| (\nabla^2 + 1) \phi \right|^2 + 
\frac{\tau}{2} \phi^2 + h(\phi) \right]\mathrm{d} \mathbf{r} ,
\end{equation}
where $|\Omega|$ is the volume of $\Omega$.

The mass conservation \eqref{eq:lbcons} implies that the order parameter $\phi$ should be constrained in $\mathcal{M}$,
\begin{equation}\label{eq:constraint}
\phi \in \mathcal{M} = \left\{  \psi \in \mathcal{L}^1_{\text{per}}(\Omega) \middle|\;\int_\Omega  \psi \; \mathrm{d} \mathbf{r} = 0\right\},
\end{equation}
and $\mathcal{P}$ is defined as a projection operator from $\mathcal{L}^1_{\text{per}}(\Omega)$ onto $\mathcal{M}$ as \begin{equation*}
    \mathcal{P}\psi = \psi-\frac{1}{|\Omega|}\int_\Omega\psi\;\mathrm{d}\mathbf{r}.\nonumber
\end{equation*}
For $\phi,v\in\mathcal{M}$, the negative gradient and negative Hessian of the LB free-energy density \eqref{eq:lbwomega} are defined as
\begin{equation}\label{eqn:fj}
    \begin{aligned}
        F(\phi)&=-Df(\phi)=-\mathcal{P}\left(\mathcal{L}\phi-g(\phi)\right),\\
        J(\phi)v&=-D^2f(\phi)v=-\mathcal{P}\left(\mathcal{L}v - \gamma\phi v + \frac{1}{2}\phi^2 v \right),
    \end{aligned}
\end{equation}
where the inner product is taken as
\begin{equation}\label{eqn:innerproduct}
    \langle\phi,\psi\rangle = \dfrac{1}{|\Omega|} \int_\Omega\phi\psi\;\mathrm{d}\mathbf{r}.
\end{equation}
It should be emphasized that for different states, the computational domain $\Omega$ can be different.
In our computations of phase transition, we will clearly specify the domain $\Omega$ for different cases, and thus $\Omega$ is omitted in the notations.

\section{Methods}
\label{method}

\subsection{Metastable states and phase diagram}
For each parameter $(\tau, \gamma)$, the stable and metastable states are the local minima of the LB free-energy density \eqref{eq:lbwomega}, which satisfy the Euler--Lagrange equation $F(\phi) = 0$.
The gradient flow
\begin{equation}\label{eq:GF}
\dfrac{\mathrm{d} \phi}{\mathrm{d} t} = F(\phi) = -\mathcal{P}\left(\mathcal{L}\phi-g(\phi)\right),
\end{equation}
which involves fourth-order derivatives, can converge to local minima from proper initial guesses. 
For the temporal discretization, a semi-implicit scheme 
\begin{equation}\label{eq:semi}
\frac{\phi^{(n+1)} - \phi^{(n)}}{\Delta t} = - \mathcal{L} \phi^{(n+1)} 
+ \mathcal{P}g(\phi^{(n)}),
\end{equation}
enables successful computation of local minima with large time step sizes using a simple formulation \cite{semipseudo}. 
Numerous fast algorithms have also been developed for the numerical identification of local minima, such as the convex splitting methods \cite{conspl}, scalar auxiliary variable algorithms \cite{sav2}, and adaptive accelerated Bregman proximal gradient (AA-BPG) methods \cite{AABPG1}.

Phase diagrams serve as quantitative visual representations that characterize the stable state of the system under varying thermodynamic conditions, specifically as functions of $(\tau, \gamma)$ in the LB model. 
Because it is hard to directly compute the global minimum of the free-energy density $f$, one can determine the stable phase by comparing the optimal free energies of multiple candidate structures, and the phase with the lowest energy corresponds to the stable state in the phase diagram. 
In the LB model, the candidate structures are as follows: the disordered phase ($\phi=0$), HEX, LAM, BCC, FCC, DG, and two Frank---Kasper phases A15 and $\sigma$.
The free energies of the equilibrium states of these candidate structures are compared to produce a phase diagram in the $\gamma$-$\tau$ plane.

For each ordered structure, the free-energy density $f$ also depends on the size of the computational domain $\Omega$.
Therefore, the computational domain should be adjusted adaptively to obtain the accurate value of the free-energy density $f$.
The optimization process of domain sizes can be summarized as follows, and the details can be found in \cite{copolymertheo2}.
From an initial computational domain, the free-energy density $f$ is minimized by numerically solving \eqref{eq:GF}, obtaining the order parameter $\phi$ at the grid points. 
Then, the free-energy density $f$ is minimized with respect to the three edge lengths of the computational domain \cite{adaptbox}, while the grid is proportionally scaled and the order parameter values at the grid points remain unchanged \cite{cam2,diagram}. 
The free-energy density $f$ is iteratively optimized with respect to the order parameter on grid points and the size of the computational domain in an alternating manner, until the free-energy density converges with a tolerance of $10^{-9}$.

For each $(\tau, \gamma)$, multiple local minima could exist in the LB model. 
For each ordered phase $s$ in $S = \{ \text{HEX}, \text{LAM}, \text{BCC}, \text{FCC}, \text{DG}, \sigma, \text{A15} \}$, the initial configuration $\phi_{s}^{(0)}$ is properly constructed based on reciprocal vectors \cite{diagram, MCMS}. 
From the initial configuration of each phase, the above optimization process of domain sizes is implemented.
It warrants additional attention that the converged solution should maintain the same symmetry as the initial configuration, which is examined by the coefficients in the reciprocal space that exceed a prescribed tolerance.
Otherwise, the converged solution actually represents a different phase and should be excluded from this phase. 
By comparing the free-energy density of the converged solution for each phase, the stable state for the current parameter is identified. 

\subsection{Landau--Brazovskii saddle dynamics}
For the same $(\tau, \gamma)$, multiple local minima could exist in the LB model.
The coexistence of multiple (meta)stable states under identical parameter conditions enables the exploration of transition pathways between these phases.
Notably, it has been observed experimentally and theoretically that the ordered phases can be related epitaxially \cite{copolymer2}. 

To investigate transition pathways, the parameters $(\tau,\gamma)$ should be properly selected based on the phase diagram.
A large energy difference between two phases is advantageous, as it results in a relatively small critical nucleus with reduced boundary effects.
Additionally, limiting the presence of other local minima mitigates their impact on the overall complexity of the energy landscape.  
Most importantly, a suitable computational domain $\Omega$ must be selected to stabilize the coexisting metastable and stable states in order to identify the transition state and the transition pathway in numerical implementations.

With an appropriately-chosen computational domain $\Omega$, we develop an LBSD method based on the high-index saddle dynamics method \cite{HiOSD,yin2021searching} to explore saddle points on the energy landscape, including transition states connecting the metastable and stable states. 
The LBSD for index-$k$ saddle points ($k$-LBSD) is,
\begin{widetext}
\begin{equation}\label{eq:lbhisd}
\left\{
\begin{aligned}
\dfrac{\mathrm{d} \phi}{\mathrm{d}t} =&
-\left(I -\sum_{j=1}^k 2v_jv_j^\top \right)
\mathcal{P}\left(\mathcal{L}\phi-g(\phi)\right),\\
\dfrac{\mathrm{d}v_i}{\mathrm{d}t}=&-\left( I-v_iv_i^\top-\sum_{j=1}^{i-1}2v_jv_j^\top\right)\mathcal{P}\left(\mathcal{L}v_i - \gamma\phi v_i + \frac{1}{2}\phi^2 v_i \right), \qquad i = 1,\cdots,k.\\
\end{aligned}
\right.
\end{equation}
\end{widetext}
The initial condition should satisfy that $\phi^{(0)}$ and $v_i^{(0)}$ lie in the constrained space $\mathcal{M}$.

A stationary state $\phi^*\in \mathcal{M}$, i.e., $Df(\phi^*)=0$, is called an index-$k$ saddle point $\phi^*$, if the Hessian $D^2 f(\phi^*)$ has exactly $k$ negative eigenvalues \cite{milnormorse}.
Then, the index-$k$ saddle point $\phi^*$ is a local maximum along the eigenvectors corresponding to these $k$ negative eigenvalues of the Hessian, and a local minimum along those corresponding to positive eigenvalues.
The $\phi$-dynamics in the $k$-LBSD utilizes this minimax structure of saddle points by implementing gradient ascent along $k$ orthonormal vectors $v_1,\cdots,v_k$ and gradient descent along other orthogonal directions. 
Accordingly, $v_1,\cdots,v_k$ directions should approximate the eigenvectors corresponding to the smallest $k$ eigenvalues of the current Hessian $D^2 f(\phi)$, and the $v_i$-dynamics in the $k$-LBSD solves this partial eigenvalue problem.

The LB model is highly ill-conditioned because of the fourth-order spatial derivatives. 
Therefore, in a similar manner to \eqref{eq:semi}, a semi-implicit scheme 
\begin{equation}\label{eq:time}
\begin{aligned}
    &\dfrac{\phi^{(n+1)}-\phi^{(n)}}{\beta^{(n)}} =  
    -\mathcal{L} \phi^{(n+1)}+ \mathcal{P}g(\phi^{(n)})\\
    &\qquad\qquad\qquad\qquad  -\sum_{j=1}^k 2v_j^{(n)}{v_j^{(n)}}^\top F(\phi^{(n)}),
    \end{aligned}
\end{equation}
is applied to discretize the $\phi$-dynamics of LBSD.
Note that the gradient flow method \eqref{eq:GF} and its semi-implicit scheme \eqref{eq:semi} can be regarded as the $k=0$ case of the LBSD \eqref{eq:lbhisd} and its semi-implicit scheme \eqref{eq:time}.

The step size $\beta^{(n)}$ is determined according to the Barzilai--Borwein gradient method \cite{bb,HiOSD}. 
Denote by $r^{(n)}$ the right-hand side of the $\phi$-dynamics in \eqref{eq:lbhisd} at the $n$th iteration,
and $\Delta r^{(n)}= r^{(n)}-r^{(n-1)}$, $\Delta \phi^{(n)} = \phi^{(n)}-\phi^{(n-1)}$.
The step size in \eqref{eq:time} is determined as
\begin{equation*}
\beta^{(n)}=\min\left\{\beta_0, \left|
\frac{\langle\Delta r^{(n)}, \Delta \phi^{(n)}\rangle}
{\langle\Delta r^{(n)}, \Delta r^{(n)}\rangle}
\right|
\right\}.
\end{equation*}
with a maximal step size $\beta_0=0.4$.

Once a transition state (an index-1 saddle point) is obtained, the MEP associated with this transition state can be constructed straightforward. 
The transition state is slightly perturbed along the eigenvector direction corresponding to the only negative eigenvalue on both sides, and the gradient flow dynamics \eqref{eq:GF} is subsequently simulated to the connected minima.
These two gradient flow trajectories form the MEP.

\subsection{Upward/downward search algorithms}\label{search}

We employ the upward/downward search algorithms based on LBSD methods to find transition states, while the degeneracy of stationary states should be especially taken into consideration in the algorithm implementation. 
When a continuous translational symmetry is broken down to a discrete translational symmetry in an ordered phase, Goldstone modes emerge, whose number equals the multiplicity of zero eigenvalues in the Hessian \cite{goldstonebrokensym,lb4}.
Consequently, the ordered phases corresponding to minima of the LB model are all degenerate: i.e., the Hessian $D^2 f$ of the system possesses zero eigenvalues.
Specifically, the multiplicity of zero eigenvalues is one for the LAM phase, two for the HEX phase, and three for other ordered phases in $S$, whereas the disordered phase has no zero eigenvalues.

The upward search algorithm is used to search for higher-index saddle points from a local minimum or a lower-index saddle point.
From an index-$k$ stationary state $\phi^*$ with $m$ zero eigenvalues, denote by $v_i^*\in \mathcal{M}$, $i = 1, \cdots, k+m+1$, the eigenvectors corresponding to the smallest $(k+m+1)$ eigenvalues of $D^2 f(\phi^*)$ in ascending order. 
To search for a higher-index saddle point, we choose a stable direction $v^*_{\mathrm{u}}$ as the perturbation direction of $\phi^*$.
Then, the upward search algorithm employs a $(k+m+1)$-LBSD simulated from $\phi^* + \varepsilon v^*_{\mathrm{u}}$ with $v^*_1, \cdots, v_{k+m}^*$ and $v^*_{\mathrm{u}}$ as the initial directions.
The perturbation direction $v^*_{\mathrm{u}}$ is typically chosen as the eigenvector corresponding to the smallest positive eigenvalue of $D^2 f(\phi^*)$, namely, $v_{k+m+1}^*$. The parameter $\varepsilon>0$ is a small constant that pushes the system away from the stationary state. 
In the LB model, the smallest positive eigenvalue for each metastable ordered state is repeated, and different choice of $v^*_{\mathrm{u}}$ may lead to different results.

From a high-index saddle point, the downward search algorithm is implemented to search for connected saddle points following its unstable directions. 
Given an index-$k$ saddle point $\phi^\star$ as the parent state with its eigenvectors $v_i^\star, i = 1, \cdots, k$ corresponding to the $i$th smallest eigenvalue of the Hessian at $D^2 f(\phi^\star)$, we choose an unstable direction $v^\star_{\mathrm{d}}$ in the space spanned by $\{v_i^\star\}_{i=1}^{k}$ as the perturbation direction.
To search for an index-$m$ ($m < k$) saddle point, the downward search algorithm employs an $m$-LBSD simulated from $\phi^\star + \varepsilon v^\star_{\mathrm{d}}$ with $m$ unstable directions orthogonal to $v^\star_{\mathrm{d}}$ as the initial directions. 
This procedure is then repeated to newly-obtained saddle points until minima are found. 
By taking different choices of $v^\star_{\mathrm{d}}$, the downward search algorithm can obtain multiple lower-index saddle points from one parent state.

In our numerical implementations, we apply the above algorithms to identify the MEP connecting the metastable state and the stable state. 
We first implement the upward search algorithm from the metastable state $\phi^*$ with $m$ zero eigenvalues.
Denote by $v_i^*$ the eigenvector corresponding to the $i$th smallest eigenvalues of $D^2 f(\phi^*)$.
An $(m+1)$-LBSD method is simulated from $\phi^* + \varepsilon v^*_{m+1}$ with $v^*_1, \cdots, v^*_{m+1}$ as the initial directions, and converges to a stationary state $\phi^{\text{new}}$. 
In most cases, $\phi^{\text{new}}$ turns out to be a high-index degenerate saddle point, for instance, an index-$k$ ($k\leq m+1$) saddle point with several zero eigenvalues. 
Denote $v^{\text{new}}_i$ as the eigenvector corresponding to the $i$th smallest eigenvalues of Hessian $D^2 f(\phi^{\text{new}})$ in ascending order.
Then, we implement the downward search algorithm from $\phi^{\text{new}}$ by simulating $k_1$-LBSD ($0<k_1<k$) with $\phi^{\text{new}} \pm \varepsilon v^{\text{new}}_k$ as the initial state and $\{v^{\text{new}}_i\}_{i=1}^{k_1}$ as the initial direction.
The perturbation direction can also be other unstable directions to obtain more results. 
By repeating this procedure to newly-found saddle points, we can locate the transition states and recover the MEP with gradient flow.

\subsection{Spatial discretizations}
\label{discretization}
Under the framework of CAM, the order parameter $\phi$ is a periodic function on the domain $\Omega$ and we apply the Fourier pseudospectral method to discretize the order parameter $\phi$.
This method has demonstrated computational efficiency in phase-field models by evaluating gradient terms in the Fourier space and nonlinear terms in the physical space \cite{cam2,PNAShisd,semipseudo}.

We discretize the cubic computational domain $\Omega=[0,L]^3$
using a uniform mesh with $N$ (an even number) grid points along each spatial coordinate.
In general, the lengths of the box domain edges may be different. 
A cubic domain is stated here for simplicity, which is consistent with our numerical experiments for transition pathways.
The order parameter $\phi$ is represented by its value $[\phi(\mathbf{r}_{ijk})]$ on the grid points, 
\begin{equation*}
\mathbf{r}_{ijk} = \frac{L}{N}(i, j, k), \quad
i,j,k=0,\cdots, N-1.\nonumber
\end{equation*}
Because of the periodicity of $\phi$ in each coordinate direction, $\phi$ can be represented as a Fourier series, which enables our implementation of the Discrete Fourier Transform (DFT) within the computational procedure.
The DFT of $[\phi(\mathbf{r}_{ijk})]$, denoted by 
$[ \hat{\phi}(\mathbf{q}_{\hat{i}\hat{j}\hat{k}})]$, is defined as
\begin{equation}\label{eq:dft}
\begin{aligned}
&\hat{\phi}(\mathbf{q}_{\hat{i}\hat{j}\hat{k}}) = 
\sum_{i,j,k=0}^{N-1} \phi(\mathbf{r}_{ijk}) \exp(- \mathrm{i} \mathbf{q}_{\hat{i}\hat{j}\hat{k}} \cdot \mathbf{r}_{ijk}),\\
&\mathbf{q}_{\hat{i}\hat{j}\hat{k}} = \frac{2\pi}{L}\left( \hat{i}, \hat{j}, \hat{k}\right), \quad
\hat{i}, \hat{j}, \hat{k} = -\frac{N}{2}, \cdots, \frac{N}{2}-1.
\end{aligned}
\end{equation}
Accordingly, $[\phi(\mathbf{r}_{ijk})]$ can also be obtained from the inverse DFT as
\begin{equation}\label{eq:invdft}
\phi(\mathbf{r}_{ijk})  = \frac{1}{N^3} 
\sum_{\hat{i},\hat{j},\hat{k}=-N/2}^{N/2-1} 
\hat{\phi}(\mathbf{q}_{\hat{i}\hat{j}\hat{k}})
\exp(\mathrm{i} \mathbf{q}_{\hat{i}\hat{j}\hat{k}} \cdot \mathbf{r}_{ijk}). 
\end{equation}

The application of DFT facilitates the conversion of differential operations in physical space into algebraic multiplications in Fourier space.
Let $\mathcal{F}_\mathbf{q}[\phi]$ denote the operator yielding the DFT component $\hat{\phi}(\mathbf{q})$ of $\phi$ corresponding to the reciprocal vector $\mathbf{q}$. 
Then, the high-order derivative term $(\nabla^2 + 1)^2 \phi^{(n+1)}$ in \eqref{eq:time} can be simply calculated in the reciprocal space as
\begin{equation}\label{eq:dfta}
\mathcal{F}_{\mathbf{q}}[(\nabla^2 + 1)^2\phi^{(n+1)}]
=(1-|\mathbf{q}|^2)^2\mathcal{F}_{\mathbf{q}}[\phi^{(n+1)}].
\end{equation}
To enforce the mass-conservation constraint \eqref{eq:constraint}, we impose $\mathcal{F}_{\mathbf{0}}[\phi] = 0$ throughout the numerical iterations.

To calculate the unstable directions $v_i$ in LBSD, the locally optimal block preconditioned conjugate gradient method \cite{lobpcg} is applied with a preconditioner \cite{PNAShisd}:
\begin{equation*}
    ((1+\nabla^2)^2+|\tau|+\epsilon)^{-1},\quad \epsilon>0, \nonumber
\end{equation*}
which can be easily calculated in the reciprocal space in a similar manner. 

\section{Results}\label{results}

\begin{figure*}[hbt]
\centering
\includegraphics[width=\linewidth]{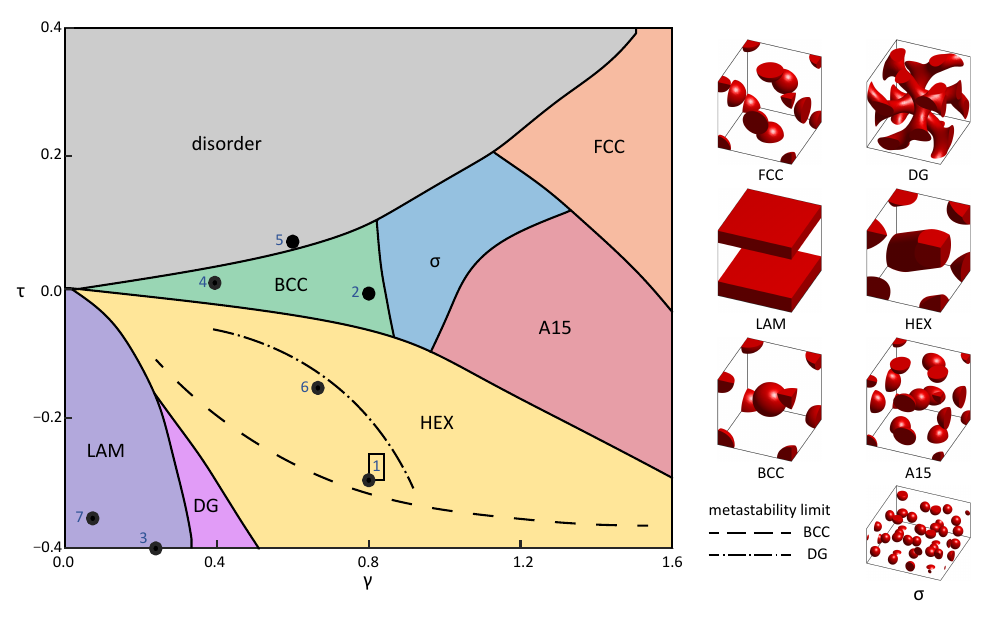}
\caption{
Phase diagram of the LB model in the $\gamma$-$\tau$ parameter space.
Colored regions separated by solid lines indicate different stable phases (disorder, FCC, DG, LAM, HEX, BCC, A15, and $\sigma$), and representative configurations of the order parameters of ordered phases are illustrated in the right panel.
Dashed lines are the metastability limit of BCC and DG.
Black dots mark parameter sets used in the following cases, with numbers labeling each one.
}
\label{fig:diagram}
\end{figure*}

\subsection{Phase diagram}
The phase diagram of the LB model in the $\gamma$-$\tau$ parameter space is constructed using the open-source software package AGPD \cite{AGPD}, which employs the AA-BPG method \cite{AABPG1} to rapidly locate different stationary states. 
The resulting phase diagram for the LB model is shown in \Cref{fig:diagram}, which is consistent with previous studies \cite{AClbmodel,diagram,dawson2021ginzburg}. 
In addition to the equilibrium phase boundaries, the phase diagram also shows some metastability limits (also known as spinodal boundaries) of the BCC and DG phases.
These boundaries represent the parameters at which a metastable state loses local stability and is no longer a local minimum.
They are included to facilitate the study of phase transitions.

\subsection{Nucleus boundary}

To identify the structure of critical nuclei across different parameters, a pointwise stable-phase-density function $\Phi_{\mathrm{s}}(\mathbf{r})$ is defined for an order parameter $\phi$ by \cite{scft3,lb3}
\begin{equation}\label{eq:critical}
\begin{aligned}
    \Phi_{\mathrm{s}}(\mathbf{r}) &= \frac{|\phi - \phi_{\mathrm{m}}|(\mathbf{r})}{|\phi - \phi_{\mathrm{m}}|(\mathbf{r}) + |\phi - \phi_{\mathrm{s}}|(\mathbf{r})},\\
|\phi_1 - \phi_2|(\mathbf{r}) &= \int w(\mathbf{r}' - \mathbf{r})[\phi_1(\mathbf{r}') - \phi_2(\mathbf{r}')]^2 \mathrm{d}\mathbf{r}',
\end{aligned}
\end{equation}
where $\phi_{\mathrm{m}}$ and $\phi_{\mathrm{s}}$ are the initial metastable state and the final stable state, respectively.
The weight function $w$ is a normalized mollifier of the form $w(\mathbf{r})\propto \exp\left(\frac{R^2}{|\mathbf{r}|^2-R^2}\right)$.
The nucleus boundary, shown as white lines in all figures, is taken to be the level set $\Phi_{\mathrm{s}}=\Phi_{\mathrm{s}}^*$, with $\Phi_{\mathrm{s}}^*$ fixed at $0.1$ in all cases. 
The radius $R$ of the mollifier is chosen so that the resulting phase density is smooth yet still reflects local differences between the new and old phases. 
In the numerical implementations, $R$ is set as $2\sqrt{2}\pi$ in cases 1--5, and $\sqrt{6}\pi$ in cases 6--7 where DG is involved.

\subsection*{Case 1: BCC--HEX transition}
First, we consider the phase transition from the BCC to the HEX phase.
For most parameters, both phases approximately share an optimal domain size of $[0, 2\sqrt{2}\pi]^3$, where the size of the reciprocal cell is $1\times1\times1$. 
Along the $[111]$ direction, the cylinders of HEX and the spheres of BCC are aligned. 
In order to accommodate the critical nucleus, the computational domain size is set as $L=24\sqrt{2}\pi$ that corresponds to $12\times12\times12$ unit cells. 
The parameter is set as $\tau=-0.3$, $\gamma=0.8$, where HEX is the stable phase with $f=-0.2004$, and BCC is the metastable phase with $f=-0.1660$. 
A mesh of $N=128$ grid points is applied along each spatial coordinate for all numerical cases for transition pathways.

\begin{figure}[htbp]
\includegraphics[width=\linewidth]{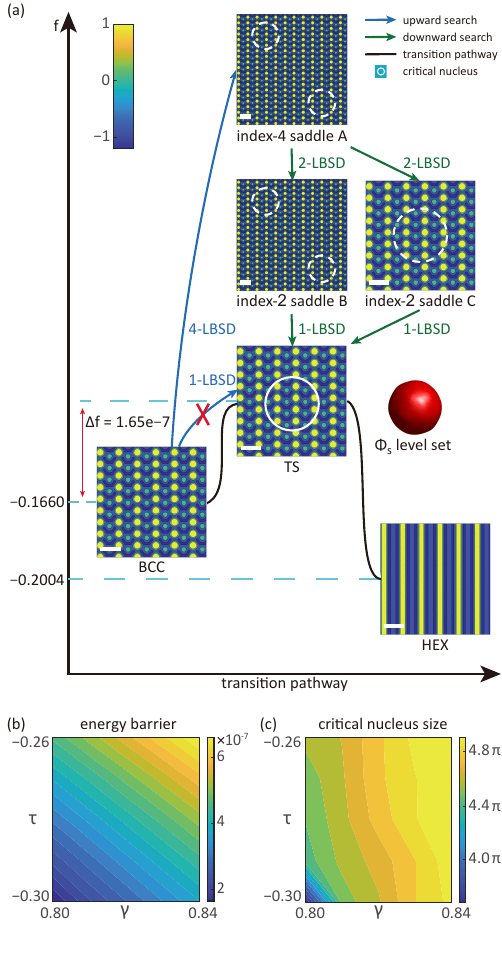}
\caption{ \label{fig:bcc2hex_030080} 
(a) Transition pathway from BCC to HEX at $ \tau = -0.3$, $ \gamma = 0.8$. 
In all cases, the ordered phases and saddle points are three-dimensional structures within $\Omega=[0,L]^3$, while we present local two-dimensional slice planes of the domain for clarity of visualization, specifically the $[10\bar{1}]$--$[111]$ plane in this figure.
For all figures, TS denotes the transition state, the critical nucleus is enclosed by the white line, and scale bars represent $4\pi$. 
White dashed lines indicate nuclei other than the critical nucleus. 
To the right of TS, the nuclear boundary defined as the $\Phi_{\mathrm{s}}$ level set of $\Phi_{\mathrm{s}}(\mathbf{r}) = \Phi_{\mathrm{s}}^*$ is shown as a red surface.
(b) Energy barrier and (c) critical nucleus size $l_x$ in $\tau\in[-0.3,-0.26]$, $\gamma\in [0.8,0.84]$.
}
\end{figure}

For the BCC phase, the Hessian matrix exhibits three zero eigenvalues, corresponding to the translational invariance of the system, and the smallest positive eigenvalue occurs with a multiplicity of 12. 
Because of the presence of three zero eigenvalues, $k$-LBSD for $k=1,2,3$ from the BCC phase cannot climb out of the basin and fails to identify a saddle point.
Therefore, we implement the upward search algorithm using 4-LBSD, successfully locating an index-4 saddle point A, from which the system can relax to the initial BCC phase and the target HEX phase through gradient flow trajectories under small perturbations. 
In general, to climb out of a local minimum with $m$ zero eigenvalues, an upward search algorithm is implemented using $k$-LBSD with $k=m+1$, and we adopt this choice in the following cases.
In this case, this upward search using 4-LBSD successfully identifies a stationary state that facilitates escape from the BCC phase with an energy increase of $3.27\times 10^{-7}$, and consequently the subsequent downward search can be implemented from this saddle point in order to discover a transition state that constitutes the transition pathway from BCC to HEX.
By comparing to the initial BCC phase using the function $\Phi_{\mathrm{s}}(\mathbf{r})$, it is discovered that this index-4 saddle point A exhibits differences from BCC that are localized within two distinct nuclei, as specifically marked by the white dashed contour representing $\Phi_{\mathrm{s}}(\mathbf{r}) = \Phi_{\mathrm{s}}^*$ in \Cref{fig:bcc2hex_030080}. 
As shown in \Cref{tab:bcc2hex030}, the saddle point A possesses four negative eigenvalues.
Two of these are on the order of $10^{-2}$ and their corresponding eigenvectors lead to the expansion and contraction of the nuclei; the other two are on the order of $10^{-4}$, with the corresponding eigenvectors leading to the translation of the nuclei.

\begin{table}[htb]
\caption{\label{tab:bcc2hex030}Hessian eigenvalues and free-energy differences for the stationary states at $\tau=-0.3, \gamma=0.8$.}
\centering
\begin{tabular}{l c c}
\hline\hline
Index   & $\Delta f$\footnote{Relative to the initial state.} & Smallest eigenvalues\footnote{For minima, only the zero and smallest positive eigenvalues are listed; for saddle points, only nonpositive eigenvalues are listed.} \\ \hline
0 (BCC) & $0$ & $0 ~(\times 3)$\footnote{Represents repeated eigenvalues; the number denotes their multiplicity; eigenvalues with absolute values smaller than $10^{-8}$ are treated as zero. The same applies below.}; $3.13 \times 10^{-3} ~(\times 12)$ \\
4 (A)   & $\phantom{-}3.27 \times 10^{-7}$ & $-2.61 \times 10^{-2} ~(\times 2)$; $-7.02 \times 10^{-4}$; \\
        &                       & $-5.74 \times 10^{-4}$; $0 ~(\times 3)$ \\
2 (B)   & $\phantom{-}3.26 \times 10^{-7}$ & $-2.73 \times 10^{-2} ~(\times 2)$; $0 ~(\times 3)$  \\
2 (C)   & $\phantom{-}1.66 \times 10^{-7}$ & $-2.67 \times 10^{-2}$; $-7.69 \times 10^{-4}$; $0 ~(\times 3)$  \\
1 (TS)  & $\phantom{-}1.65 \times 10^{-7}$ & $-2.83 \times 10^{-2}$; $0 ~(\times 3)$  \\
0 (HEX) & $-3.44 \times 10^{-2}$ & $0 ~(\times 2)$; $2.40 \times 10^{-4} ~(\times 4)$ \\ \hline\hline
\end{tabular}
\end{table}

Then, we implement the downward search algorithm using 2-LBSD by perturbing A along the eigenvectors corresponding to the latter two eigenvalues, which locates an index-2 saddle point B. 
This saddle point B also possesses two spherical nuclei, while the position of the nuclei is slightly different from the saddle point A. 
The center of each nucleus of A is located midway between two adjacent BCC spheres along the [111] direction, while the center of each nucleus of B is located at the center of one BCC sphere.
Consequently, the free-energy density of B is lower than that of A by only $1.22\times 10^{-9}$.
The saddle point B possesses two negative eigenvalues on the order of $10^{-2}$, whose corresponding eigenvectors lead to the expansion and contraction of the nuclei.

By implementing the downward search algorithm using 1-LBSD starting from B, we finally obtain an index-1 saddle point TS, which is the transition state that connects the BCC and HEX states. 
The transition state mainly resembles BCC except for a critical nucleus inside, which essentially corresponds to one nucleus of B, and the nucleus boundary calculated by $\Phi_{\mathrm{s}}(\mathbf{r}) = \Phi_{\mathrm{s}}^*$ in \eqref{eq:critical} is shown with a white line in \Cref{fig:bcc2hex_030080}(a).
The eigenvector corresponding to its only negative eigenvalue leads to the expansion and contraction of the nucleus. 
The energy barrier $1.65\times 10^{-7}$ is approximately half of the energy increase of B.

As another approach to TS, we can also eliminate one nucleus first from the saddle point A.
A downward search algorithm can be implemented using 2-LBSD by perturbing along a different unstable direction to locate a new index-2 saddle point C. 
This saddle point C possesses one nucleus that essentially corresponds to a nucleus of A, and the two negative eigenvalues of C are on the order of $10^{-2}$ and $10^{-4}$, respectively, with the corresponding eigenvectors of respectively similar effects.
Then, the saddle point TS can also be obtained by downward search from C accordingly, with an energy decrease of $6.13\times 10^{-10}$.

From the above saddle points, a clear correlation can be observed between its Hessian eigenpairs and nuclei.
At a given saddle point, a critical nucleus contributes to a negative eigenvalue, with the corresponding eigenvector governing its expansion or contraction. 
When this nucleus is displaced to a higher-energy position, an additional negative eigenvalue would appear, with the corresponding eigenvector leading to its translation.  
However, the contribution from the nucleus itself induces a substantially larger energy increase, yielding an eigenvalue of greater magnitude. 
For a high-index saddle point, the negative eigenvalues of the Hessian directly reflect information about the associated nuclei. 
On the other hand, the downward search from a high-index saddle point eliminates all other unstable directions, leaving only a single critical nucleus at its energy-minimized position.

Several smallest Hessian eigenvalues and free-energy differences for these aforementioned stationary states are listed in \Cref{tab:bcc2hex030}. 
\Cref{fig:bcc2hex_030080}(a) illustrates the upward (with a blue arrow) and downward search (with green arrows) procedure and the transition pathway (with black curves) from BCC to HEX.
Besides this downward search approach of sequentially utilizing 2-LBSD and 1-LBSD, it should be noted that an alternative strategy can also accurately locate the same transition state from the saddle point A. 
For example, either a step-by-step strategy of sequentially applying 3-LBSD, 2-LBSD, and 1-LBSD, or a straightforward strategy of directly applying 1-LBSD from A, can locate TS successfully.

As previously mentioned, the smallest positive eigenvalue of the Hessian at the BCC phase has a multiplicity of 12, which provides extensive flexibility during upward search, as the initial perturbation can be chosen as any direction in the corresponding 12-dimensional eigenspace. 
In principle, different perturbation directions may lead to different saddle points, potentially identifying distinct transition states.
To investigate this issue, we perform multiple upward searches from different perturbation directions to locate distinct transition states. 
A comparison of the results shows that although different transition states are obtained from different perturbation directions, the differences in the energy barriers and negative eigenvalues of them are less than $10^{-12}$ and $10^{-5}$, respectively. 
Furthermore, the relative positions of these critical nuclei relative to the BCC spheres are nearly indistinguishable. 
Therefore, we conclude that, within numerical tolerance, these transition states effectively represent the same unique transition state.

Within a small parameter range, the BCC phase persists as metastable and the HEX phase as stable, while their free-energy densities vary accordingly.
Therefore, we investigate how the energy barrier and the size of the critical nucleus respond to changes in the free-energy difference.
Although the aforementioned method can compute transition states across varying parameters, we employ a simplified approach to improve computational efficiency.
The previously-determined transition state $\phi_0$, while not exact under new parameters, serves as a reliable initial guesses for parameters near $\tau=-0.3, \gamma=0.8$. 
Accordingly, we apply the 1-LBSD method starting from $\phi_0$ at the new parameter sets and verify that the resulting transition state indeed connects the BCC and HEX phases. 
This homotopy continuation approach is then carried out step by step for the parameters in the ranges $\tau\in[-0.3,-0.26]$, $\gamma\in[0.8, 0.84]$.

The shape of critical nuclei can be characterized as the ratio of the lengths along the three coordinates, $l_x$, $l_y$ and $l_z$, based on the nucleus boundary. 
For these critical nuclei, ${l_y}/{l_x}$ and ${l_y}/{l_z}$ are close to 1, indicating that they are spherical, so the nucleus size can be represented by $l_x$.
To visualize each critical nucleus, we first determine its position and center according to the nucleus boundary, and the slice plane should pass through the center to offer an effective visualization of the nuclear structure.

Figures \ref{fig:bcc2hex_030080}(b) and \ref{fig:bcc2hex_030080}(c) illustrate the variation of the energy barrier and critical nucleus size across the parameter space. 
As $\tau$ or $\gamma$ increases, both quantities increase, indicating that the transition from BCC to HEX becomes progressively more difficult.
This trend is consistent with the system approaching the BCC--HEX phase boundary, as shown in \Cref{fig:diagram}. 
The decreasing free-energy difference between the two phases near this boundary impedes this transition, in agreement with our computational results.

In the above simulations, the BCC and HEX phases correspond to distinct branches of local minima in the LB free-energy landscape, and phase transitions between them are characterized by transition pathways.
Along these pathways, the critical nucleus, formed by elongated BCC spheres with partial connectivity, appears as a localized region within the BCC background.
As the system evolves past the saddle point, this region grows and eventually spreads throughout the entire domain, leading to the formation of the HEX phase.
No continuous global morphological evolution from BCC to HEX is observed in our present simulations, consistent with the typical first-order nature of this transition in the LB model.

\begin{figure*}[hbtp]
\includegraphics[width=\linewidth]{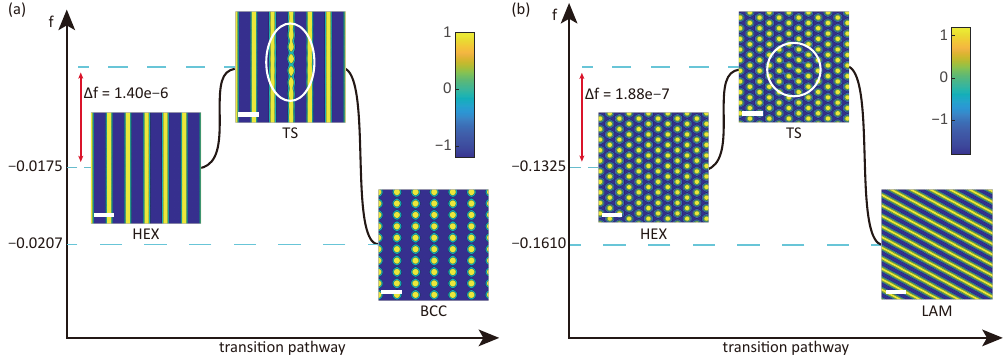}
\caption{\label{fig:combine_hbhl}
(a) Transition pathway from HEX to BCC at $\tau = -0.008$, $\gamma = 0.8$. 
The slice plane is $[10\bar{1}]$--[111]. 
(b) Transition pathway from HEX to LAM at $ \tau = -0.4$, $ \gamma = 0.22$.  
The slice plane is $[10\bar{1}]$--$[1\bar{2}1]$.}
\end{figure*}

\subsection*{Case 2: HEX--BCC transition}
The transition pathway from HEX to BCC can be obtained in a similar manner with the same $L=24\sqrt{2}\pi$. 
The parameter is set as $\tau=-0.008$, $\gamma=0.8$, where BCC is the stable phase with $f=-0.0207$, and HEX is the metastable phase with $f=-0.0175$.

The metastable HEX phase exhibits two zero eigenvalues, and the smallest positive eigenvalue occurs with multiplicity 4.
An upward search using 3-LBSD from HEX successfully locates an index-3 saddle point and a subsequent downward search using 1-LBSD directly yields a transition state connecting the HEX and BCC phases. 
\Cref{tab:hex2bcc} presents some Hessian eigenvalues and free-energy differences for these stationary points. 
The transition pathway is illustrated in \Cref{fig:combine_hbhl}(a) with an energy barrier of $1.40\times10^{-6}$.
The transition state structurally resembles the HEX phase and features a critical nucleus.

\begin{table}[htb]
\caption{\label{tab:hex2bcc}Hessian eigenvalues and free-energy differences for some stationary states at $\tau=-0.008, \gamma=0.8$.}
\centering
\begin{tabular}{l c c}
\hline\hline
Index  & $\Delta f$ & Smallest eigenvalues  \\ \hline
0 (HEX) & $0$ & $0 ~(\times 2)$; $2.62 \times 10^{-4} ~(\times 4)$  \\
3 & $\phantom{-}1.62 \times 10^{-6}$ & $-1.34 \times 10^{-2}$;  $-9.23 \times 10^{-3}$; \\
  &                       & $-4.40 \times 10^{-4}$; $0 ~(\times 3)$ \\
1 & $\phantom{-}1.40 \times 10^{-6}$ & $-1.03 \times 10^{-2}$; $0 ~(\times 3)$  \\
0 (BCC) & $-3.23 \times 10^{-3}$ & $0 ~(\times 3)$; $1.54 \times 10^{-3} ~(\times 12)$ \\ \hline\hline
\end{tabular}
\end{table}

Unlike the spherical critical nucleus in the previous case, the nucleus boundary defined via \eqref{eq:critical} in this case adopts an ellipsoidal geometry with its major axis aligned along the [111] direction---the orientation of the cylinders---as shown in \Cref{fig:combine_hbhl}(a). 
Define $l_{\parallel}$ and $l_\perp$ as the lengths along and perpendicular to the major axis ([111] direction), respectively.
Further calculations show that the aspect ratio of the critical nucleus is $l_\perp / l_\parallel \approx 1:4$.

\subsection*{Case 3: HEX--LAM transition}
In this case, we consider the transition between the HEX and LAM phases, where the cylinders of HEX are aligned along the [111] direction, and the planes of LAM are normal to the $[01\bar{1}]$ direction. 
The HEX phase is a two-dimensional structure essentially and is invariant along the [111] direction, while the LAM phase is a one-dimensional structure.
Under this configuration, a unit cell $[0, 2\sqrt{2}\pi]^3$ can serve as an optimal domain size for both phases, and $L$ is set as $28\sqrt{2}\pi$. 
The parameter is set as $\tau=-0.4, \gamma=0.22$, where LAM is the stable phase with $f= -0.1610$, and HEX is the metastable phase with $f=-0.1325$.

An upward search using 3-LBSD from HEX locates an index-4 saddle point with a negative eigenvalue close to 0. 
As this saddle point, the presence of three negative eigenvalues on the order of $10^{-2}$ is consistent with the existence of three nuclei.
Then, a downward search using 1-LBSD is applied to this saddle point, locating the transition state connecting the HEX and LAM phases with an energy barrier of $1.88\times10^{-7}$. 
The transition state is structurally analogous to the HEX phase, characterized only by a critical nucleus.
We tabulate some information for these stationary states in \Cref{tab:hex2lam}, and the corresponding transition pathway is shown in \Cref{fig:combine_hbhl}(b).
The critical nucleus exhibits an ellipsoidal morphology, analogous to that in case 2, but with its major axis oriented along the $[1\bar{1}0]$ direction, which is perpendicular to the orientation of the cylinders.
The aspect ratio of the critical nucleus is found to be $l_\perp / l_\parallel \approx 1:3$, with $l_\parallel$ and $l_\perp$ defined according to the major axis.
Although HEX and LAM phases can also be obtained in a two-dimensional LB model, the resulting critical nucleus differs fundamentally from this three-dimensional result and does not retain the local three-dimensional structure.

\begin{table}[htb]
\caption{\label{tab:hex2lam}Hessian eigenvalues and free-energy differences for some stationary states at $\tau=-0.4, \gamma=0.22$.}
\centering
\begin{tabular}{l c c}
\hline\hline
Index  & $\Delta f$ &  Smallest eigenvalues \\ \hline
0 (HEX) & $0$ & $0 ~(\times 2)$; $1.60 \times 10^{-4} ~(\times 4)$ \\
4      & $\phantom{-}5.68 \times 10^{-7}$ & $-3.91 \times 10^{-2}$; $-3.90 \times 10^{-2} ~(\times 2)$; \\
       &                       & $-3.63 \times 10^{-5}$; $0 ~(\times 3)$  \\
1      & $\phantom{-}1.88 \times 10^{-7}$ & $-3.90 \times 10^{-2}$; $0 ~(\times 3)$  \\
0 (LAM) & $-2.85 \times 10^{-2}$ & $0$; $1.47 \times 10^{-5} ~(\times 2)$ \\ \hline\hline
\end{tabular}
\end{table}

\begin{figure*}
\includegraphics[width=\linewidth]{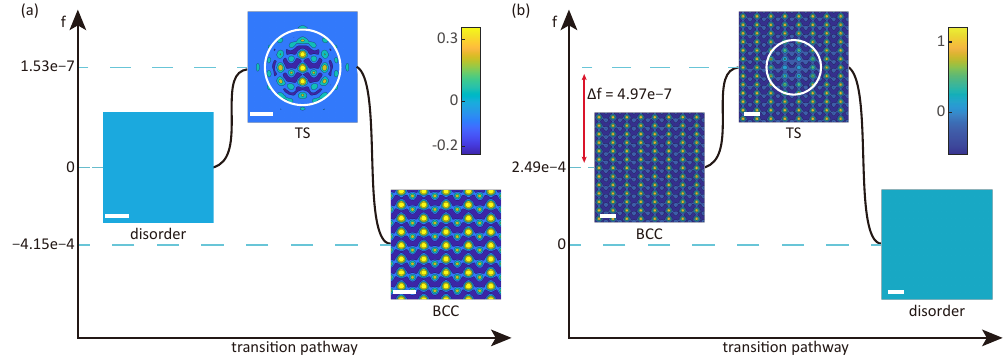}
\caption{\label{fig:combine_bddb}
(a) Transition pathway from disorder to BCC at $ \tau = 0.01$, $ \gamma = 0.4$. The slice plane is $[10\bar{1}]$--$[111]$. 
(b) Transition pathway from BCC to disorder at $ \tau = 0.05 $, $ \gamma = 0.6$. The slice plane is $[10\bar{1}]$--$[111]$.}
\end{figure*}

\subsection*{Case 4: disorder--BCC transition}
In this case, we investigate the nucleation process of the BCC phase from disordered phase. 
The disordered phase $\phi=0$ is locally stable with $f=0$ when $\tau>0$. 
Because $[0,2\sqrt{2}\pi]^3$ is an optimal domain size of the BCC phase, $L$ is set as $20\sqrt{2}\pi$. 
The parameters are set as $\tau=0.01, \gamma=0.4$ where BCC is the stable phase with $f=-4.15\times 10^{-4}$. 

For the disordered phase, the Hessian matrix exhibits no zero eigenvalues, but possesses an eigenvalue $\tau$ with high multiplicity. 
Theoretically, any function of the form $v(\mathbf{r}) = \exp(\mathrm{i} \mathbf{k} \cdot \mathbf{r})$ with $|\mathbf{k}| = 1$ serves as an eigenfunction corresponding to this eigenvalue.
More importantly, for the upward search using 1-LBSD, different perturbation directions lead to different saddle points and thus different local minima through gradient flow.
By choosing the perturbation direction as the BCC phase in $[0,L/2]^3$, we can locate the desired transition state that connects BCC, in which the critical nucleus exhibits a spherical morphology with a BCC-ordered internal structure.
Numerical experiments demonstrate that randomly selecting an eigenvector associated with the eigenvalue $\tau$ as the perturbation direction generally fails to find the desired transition state. 
Instead, the obtained index-1 saddle point is connected to a BCC-like state with a higher energy and a different orientation.

\subsection*{Case 5: BCC--disorder transition}
In the case of transition from BCC to the disordered phase, $L$ is set as $32\sqrt{2}\pi$.
The parameters are set as $\tau=0.05, \gamma=0.6$ where BCC is the metastable phase with $f= 2.49\times10^{-4}$.

The Hessian of the initial BCC phase has three zero eigenvalues, so an upward search algorithm using 4-LBSD is implemented. 
Different from case 1, this search yields an index-2 saddle point and the other two eigenvalues are zero.
A subsequent downward search using 1-LBSD locates an index-1 saddle point that is the transition state connecting the BCC and disorder phases. 
The transition state structurally resembles the BCC phase and contains a spherical critical nucleus.
The corresponding transition pathway is depicted in \Cref{fig:combine_bddb}(b), with an energy barrier of $4.97\times10^{-7}$.

\begin{figure*}
\includegraphics[width=\linewidth]{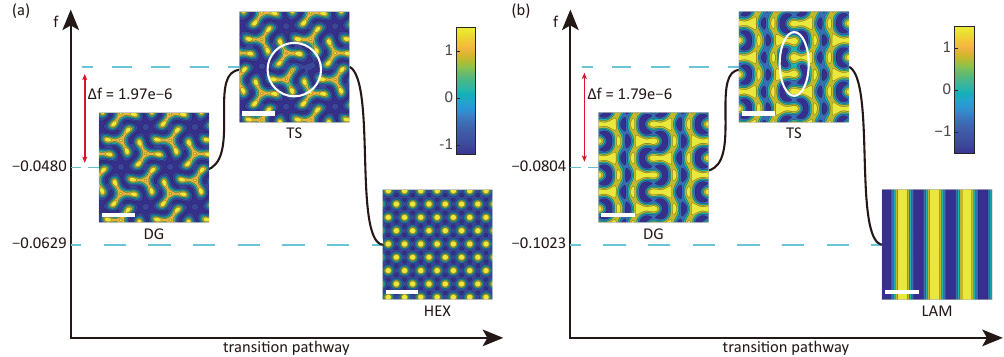}
\caption{\label{fig:slice_dh}
(a) Transition pathway from DG to HEX at $ \tau = -0.14 $, $ \gamma = 0.7$.  The slice plane is $[10\bar{1}]$--$[1\bar{2}1]$. 
(b) Transition pathway from DG to LAM at $ \tau = -0.32 $, $ \gamma = 0.08 $. The slice plane is $[10\bar{1}]$--$[111]$.}
\end{figure*}

\subsection*{Case 6: DG--HEX transition}
The optimal domain size of the DG phase is $[0, 2\sqrt{6}\pi]^3$, while that of the HEX phase is $[0, 2\sqrt{2}\pi]^3$ with $[111]$ as the cylinder direction.
By a 30-degree rotation of the HEX phase around the cylinder direction, the unit cell $[0, 2\sqrt{6}\pi]^3$ is simultaneously optimal for the HEX phase.
In this unit cell, the cylinders of HEX remain aligned along the [111] direction, while the arrangement of cylinders is different.
Therefore, in this case, the computational domain size is set as $L=8\sqrt{6}\pi$ that corresponds to $4\times4\times4$ unit cells. 
The parameter is set as $\tau=-0.14, \gamma=0.7$ where DG is the metastable phase with $f=-0.0480$, and HEX is the stable phase with $f=-0.0629$. 

The Hessian of the metastable DG phase has three zero eigenvalues, so an upward search algorithm using 4-LBSD is implemented by perturbing along the eigenvector associated with the smallest positive eigenvalue.
This upward search identifies an index-2 saddle point with two nuclei inside. 
Then, a downward search employed using 1-LBSD locates an index-1 saddle point that connects DG and HEX with an energy barrier of $2.0\times 10^{-6}$.
Several smallest Hessian eigenvalues for DG, HEX, and the aforementioned saddle points are listed in \Cref{tab:dg2hex}, and the transition pathway from DG to HEX is illustrated in \Cref{fig:slice_dh}(a). 
The nucleus boundary defined via \eqref{eq:critical} exhibits an ellipsoidal morphology with its major axis along the $[001]$ direction, and the aspect ratio of the critical nucleus is $l_\perp / l_\parallel \approx 1:1.2$.

\begin{table}[htbp]
\caption{\label{tab:dg2hex}Hessian eigenvalues and free-energy differences for some stationary states at $\tau=-0.14, \gamma=0.7$.}
\centering
\begin{tabular}{l c c}
\hline\hline
Index  & $\Delta f$     &  Smallest eigenvalues \\ \hline
0 (DG)  & $0$ & $0 ~(\times 3)$; $7.78 \times 10^{-3} ~(\times 12)$ \\
2      & $\phantom{-}3.96 \times 10^{-6}$ & $-6.58 \times 10^{-3}$; $-6.32 \times 10^{-3}$; $0 ~(\times 3)$ \\
1      & $\phantom{-}1.97 \times 10^{-6}$ & $-6.48 \times 10^{-3}$; $0 ~(\times 3)$  \\
0 (HEX) & $-1.49 \times 10^{-2}$ & $0 ~(\times 2)$; $1.24 \times 10^{-3} ~(\times 4)$ \\ \hline\hline
\end{tabular}
\end{table}

\subsection*{Case 7: DG--LAM transition}
Finally, we consider the transition from the DG phase to the LAM phase.
$L=8\sqrt{6}\pi$ is also approximately optimal for the LAM phase. 
The parameters are set as $\tau=-0.32$, $\gamma=0.08$, where DG is the metastable phase with $f=-0.0804$ and LAM is the stable phase with $f=-0.1023$.

An upward search algorithm using 4-LBSD from the DG metastable state yields an index-4 saddle point, and a subsequent downward search algorithm using 1-LBSD locates an index-1 saddle point that connects DG and LAM. 
For the LAM phase identified from the transition state, the planes are normal to the $[01\bar{1}]$ direction.
Several Hessians eigenvalues and the free-energy differences of these stationary states are listed in \Cref{tab:dg2lam}, and the transition pathway is illustrated in \Cref{fig:slice_dh}(b)
with an energy barrier $1.8\times10^{-6}$. The transition state mainly resembles DG except for a local nucleus in the middle of the figure.
The critical nucleus exhibits an ellipsoidal morphology with its major axis along the $[111]$ direction, and the aspect ratio of the critical nucleus is
$l_\perp / l_\parallel \approx 1:2$.

\begin{table}[htbp]
\caption{\label{tab:dg2lam}Hessian eigenvalues and free-energy differences for some stationary states at $ \tau=-0.32, \gamma=0.08$.}
\centering
\begin{tabular}{l c c}
\hline\hline
Index & $\Delta f$  &  Smallest eigenvalues \\ \hline
0 (DG) & $0$ & $0 ~(\times 3)$; $6.67 \times 10^{-3} ~(\times 12)$ \\
4     & $\phantom{-}4.14 \times 10^{-6}$ & $-1.53 \times 10^{-2}$; $-1.51 \times 10^{-2}$; \\
      &                       & $-5.62 \times 10^{-3}$; $-5.52 \times 10^{-3}$; $0 ~(\times 3)$ \\
1     & $\phantom{-}1.79 \times 10^{-6}$ & $-1.33 \times 10^{-2}$; $0 ~(\times 3)$  \\
0 (LAM) & $-2.19 \times 10^{-2}$ & $0$; $5.49 \times 10^{-4} ~(\times 2)$ \\ \hline\hline
\end{tabular}
\end{table}

\section{Conclusions and discussions}
In this work, we study a variety of ordered phases of the three-dimensional Landau--Brazovskii (LB) free-energy model. 
We construct the phase diagram of the three-dimensional LB model in the $\gamma$-$\tau$ parameter space, which comprises eight distinct equilibrium phases.  
Guided by the phase diagram, we systematically identify the transition pathways between different phases by employing the LB saddle dynamics method, thereby shedding light on the underlying mechanisms that govern structural transformations in modulated-phase systems described by the LB model. 
In particular, we examine seven representative cases of phase transitions involving BCC, HEX, LAM, DG, and disordered phases, and identify the corresponding critical nucleus along each transition pathway.
For the BCC--HEX transition, we further investigate (1) the uniqueness of the critical nucleus and the associated energy barrier, (2) the negative eigenvalues of high-index saddle points and the corresponding eigenvectors, and (3) the dependence of the energy barrier and the critical nucleus size under different parameter conditions.
These detailed analyses provide fundamental insights into the nucleation mechanisms underlying structural transitions.
The presented method also provides a general framework applicable to other first-order phase transitions described by Landau-type phase-field models \cite{dawson2021ginzburg}.

For the LB model, different phases generally require distinct optimal domain sizes, for instance, a cubic domain with the edge length $2\sqrt{2}\pi$, $2\sqrt{3}\pi$, $2\sqrt{5}\pi$, and $2\sqrt{6}\pi$ for the BCC, FCC, A15, and DG phases, respectively.
In constructing each transition pathway, an appropriate unified domain size is selected to ensure compatibility between the two phases involved and accommodate the corresponding critical nucleus.
Different transition pathways may require different domain sizes, and it is challenging for a single computational domain to stabilize multiple phases simultaneously.
As a compromise, one may employ the crystalline approximation method \cite{cam2,jiang2025approximation} in conjunction with Diophantine approximation techniques. 
This strategy may enable the calculation of phase transitions involving more phases, for example, FCC and A15 phases, but it often requires substantially larger computational domains and leads to excessive computational costs.

An alternative strategy is to relax the fixed-domain setting in the present study and develop algorithms that allow both the order parameter and the computational domain size to evolve simultaneously.
In practice, the computational grid is fixed in terms of the number of structured grid points, while the domain size is treated as a continuous parameter.
Varying the domain size corresponds to a rescaling of spatial coordinates, allowing the system to select energetically favorable domain sizes without altering the grid topology.
This approach provides a practical way to explore optimal domain sizes during phase transitions without the additional computational cost associated with large domains, while the systematic development of such dynamically rescaled algorithms is left for future work.

In our current study, we consider the transition pathways between two phases with parameters chosen away from phase boundaries.
As shown in case 1, approaching the phase boundary increases both the energy barrier and the size of the critical nucleus, making transitions more difficult.
Near triple junctions in the phase diagram, multiple metastable phases can coexist with comparable free-energy levels, resulting in a highly complex energy landscape with closely-spaced local minima and saddle points.
While such transitions are in principle accessible within the current methodology, resolving detailed transition pathways poses significant computational challenges, including multistability and finite-domain effects.
A systematic exploration of transitions near triple junctions would be computationally demanding and is left for future investigation.

Note that the LB framework is a Landau-theory model derived under mean-field approximations for cases where a single wavelength dominates, and the free energy is expressed as Taylor expansions in powers of the order parameter.
Within this framework, complex phases with multiple characteristic length scales, such as the A15 and $\sigma$ phases, can be captured and warrant further detailed investigation.
Nevertheless, the LB model may neglect fluctuation effects or coupling between different modes, which may become crucial in some systems or phases, because such effects can alter the stability and morphology of modulated phases and potentially shift the location or nature of phase boundaries. 
Some extended phase-field crystal models, incorporating high-order derivative terms as corrections, have been proposed to more accurately capture such effects and provide refined descriptions of the complex ordering behavior in modulated-phase systems \cite{starodumov2022review}.
For certain systems such as block copolymers, future work could also employ more sophisticated theoretical approaches, such as density functional theory or self-consistent field theory, which naturally include multimode coupling and local correlation effects, to obtain a more quantitative understanding. 
However, these methods are computationally much more demanding, and the development of efficient algorithms for exploring the corresponding high-dimensional free-energy landscapes remains an open and challenging problem.

\section*{Acknowledgments}
L.Z. was supported by the National Natural Science Foundation of China (No. 12225102, T2321001, 12288101) and National Key Research and Development Program of China 2024YFA0919500.
J.Y. was supported by ``the Fundamental Research Funds for the Central Universities'' (No.~2253100016).
K.J. was supported by the NSFC grant (12171412), and The Innovative Research Group Project of National Natural Science Foundation of Hunan Province of China (2024JJ1008). 

\section*{DATA AVAILABILITY}
The data that support the findings of this article are not publicly available. The data are available from the authors upon reasonable request.

\bibliography{lbmodel}
\end{document}